\documentclass[onecolumn,useAMS,usenatbib]{mn2e}

\usepackage{graphicx}

\title[GW emission from superfluid neutron stars ]{Gravitational wave emission from rotating superfluid neutron stars}

\author[D. I. Jones]{D. I. Jones$^1$\thanks{Email: D.I.Jones@soton.ac.uk} \\
  $^1$School of Mathematics, University of Southampton, Southampton SO17 1BJ}

\usepackage[breaklinks,hyperindex]{hyperref}

\begin{document}


\pagerange{\pageref{firstpage}--\pageref{lastpage}} \pubyear{2009}

\maketitle

\label{firstpage}

\begin{abstract}
  In this paper we investigate the effect of a pinned superfluid
  component on the gravitational wave emission of a steadily rotating
  deformed neutron star.  We show that the superfluid pinning allows
  the possibility for there to be gravitational wave emission at both
  the stellar spin frequency $\Omega$ and its first harmonic,
  $2\Omega$.  This contrasts with the conventional case where there is
  no pinned superfluidity, where either only the $2\Omega$ harmonic is
  present, or else the star undergoes precession, a feature which is
  not believed to be common in the known pulsar population.  This work
  motivates the carrying out of gravitational wave searches where both
  the $\Omega$ and $2\Omega$ harmonics are searched for, even in
  targeted searches for waves from known pulsars which aren't observed
  to precess.  Observation of such a two-component signal would
  provide evidence in favour of pinned superfluidity inside the star.
\end{abstract}


\section{Introduction}

Spinning neutron stars are a source of potentially detectable
gravitational waves, and so are of interest for the current generation
of laser interferometer gravitational wave detectors
\citep{aetal09_RPP72, aetal08_25, grote08}.  A number of searches have
already been performed, so far producing upper limits on the
gravitational wave emission \citep{aetal04_PRD69, aetal05_PRL94,
  aetal05_PRD72, aetal07_PRD76a, aetal07_PRD76b, aetal08_PRD77,
  aetal08_APJL683, aetal09_PRD79, aetal09_PRL102}.  A highlight was
the result of \citet{aetal08_APJL683}, where the energy being emitted
in the gravitational wave channel by the Crab pulsar was shown to be
no more than 6\% of the total spin-down energy budget.

Most gravitational wave searches have assumed emission at a single
gravitational wave frequency.  In the case of searches targeting
pulsars of known spin frequency $\Omega$, the gravitational wave
signal has been assumed to be at frequency $2\Omega$ (see e.g.
\citet{aetal07_PRD76a}).  This is exactly the sort of emission to be
expected from a steadily spinning triaxial body, rotating about one of
the principal axes of its moment of inertia tensor \citep{st83}.  The
necessary deformation could be supported by strains in the solid
crust, or by strong internal magnetic fields \citep{ucb00, hetal08}.
More complicated gravitational wave emission occurs, with multiple
harmonics, if the spin and principal axes are misaligned \citep{zs79,
  zimm80, ja02, vdb05}.  However, such free precession would produce
modulations in the time of arrival, polarisation, and beam shape of
the radio pulsations (see \citet{ja01} and references therein).  It is
the absence of clear evidence of precession in gravitational wave
candidates that has led to most searches assuming a single
monochromatic signal.  A notable exception was the analysis for the
Crab pulsar presented by \citet{aetal08_APJL683}, where a small
frequency band around $2\Omega$ was searched, motivated in part by the
possibility of precession, and in part by the possibility of
gravitational wave emission occurring from a stellar component with
spin frequency slightly different from the radio-pulsation-producing
outer crust.

In this paper we revisit the issue of gravitational wave emission from
rotating stars, by allowing for a superfluid component within the
star.  There are strong reasons to believe that superfluids exist in
neutron stars, and that they have an effect on the stellar dynamics
\citep{saul89}.  Superfluids rotate by forming an array of vortices,
with the number of vortices per unit area being proportional to the
rotation rate.  An interaction between these vortices and the charged
component of the star is believed to result in `pinning' of these
vortices, preventing that portion of the superfluid that is pinned
from participating in the smooth spin-down of the star.  Instead, the
spin rate of the pinned superfluid remains constant, so that a spin
lag builds up between the superfluid and the pinned vortices.  When
this lag becomes sufficiently large, some mechanism is believed to
result in a catastrophic unpinning, spinning up the star's charged
component, and producing one of the well known pulsar glitches.

As was pointed out by \citet{shah77}, in addition to producing
glitches, a pinned superfluid component also has important
repercussions for the precessional dynamics of a star.  Classically,
the free precession of a rigid body is determined by the principal
moments of inertia $(I_1, I_2, I_3)$.  However, Shaham showed that a
pinned superfluid alters the motion, effectively acting as a large
gyroscope, `sewn' into the body of the star.  If the star is set into
precession, the angular momentum vector of the pinned superfluid
follows the wobbling motion of the star, and this must be included in
the solution of the Euler rigid body equations of motion.

The detailed form of the free precession that then occurs is given in
\citet{shah77}.  We will not concern ourselves with the corresponding
gravitational wave emission here.  Instead, we will confine our
attention to the allowed non-precessional motion of such a star, and
the consequent gravitational wave emission.  The key point is this: if
one allows for the possibility of the superfluid pinning along an axis
that isn't one of the principal axes of the star's moment of inertia
tensor, even the non-precessing motion is interesting: despite
rotating steadily at rate $\Omega$, there is gravitational wave
emission at both $\Omega$ and $2\Omega$.  The relative strengths are
dependent on the moments of inertia and relative orientations of the
stellar components. Basically, this motion is possible because the
moment of inertia of the pinned superfluid is likely to be many orders
of magnitude greater than the asymmetries $(I_2-I_1, I_3-I_2,
I_3-I_1)$, which (as we show below) means that, to a good
approximation, the non-precessing motion corresponds to rotation about
the superfluid pinning axis, not to rotation about any one of the
principal axes.  This means that even in the absence of any observed
precession in the radio data, a given star might possibly be radiating
at both $\Omega$ and $2\Omega$, motivating the carrying out of
gravitational wave searches that account for this.

There are two different locations where superfluid pinning may occur
in a neutron star (see e.g. \citet{saul89}).  One location is the
inner crust, the other is the core, and the physical conditions in the
two regimes are very different.  In the crust, pinning may occur
because of an interaction between the vortices and the nuclear lattice
\citep{ai75}.  In the case of small relative velocities between
superfluid and vortex, the pinning may be strong enough to prevent
motion of the vortices with respect to the lattice; an explicit model
of this was presented recently by \citet{link09}.  This crustal
superfluid is believed to make up only a few percent of the star's
total moment of inertia.  The other possible pinning location is the
core, where there is believed to be a strong interaction between the
vortices and magnetic flux tubes \citep{rude76, rude91a, rude91b}.  In
this case, the bulk of the star's moment of inertia might be a pinned
superfluid.  For the most part, the exact location and cause of the
pinning is not important for our purposes---what matters is that the
pinning exists and is maintained for periods longer than the
gravitational wave observation time.  The long separation between
glitches in young pulsars indicates that this may indeed be the case
\citep{lg98}.

The plan of this paper is as follows.  In section \ref{sect:bm} we
present our basic rigid body stellar model and derive the form of the
non-precessional solution.  In section \ref{sect:gwe} we calculate the
corresponding gravitational wave emission, count the number of
parameters required to describe the signal, and discuss the angular
pattern of the emission.  In section \ref{sect:etaes} we show that the
results obtained still apply when the assumption of rigidity is
relaxed to one of elasticity.  In section \ref{sect:lac} we discuss
what might limit the wave emission and what mechanisms might prevent
our model being realised.  In section \ref{sect:ofsm} we talk about
two other (non-superfluid) mechanisms that can lead to radiation at
multiple frequencies, and show they are likely to be less important
than the superfluid pinning mechanism.  Finally, in section
\ref{sect:discussion}, we speculate as to how the sorts of deformed
stars considered in this paper might arise in nature.

\section{Basic Model}
\label{sect:bm}

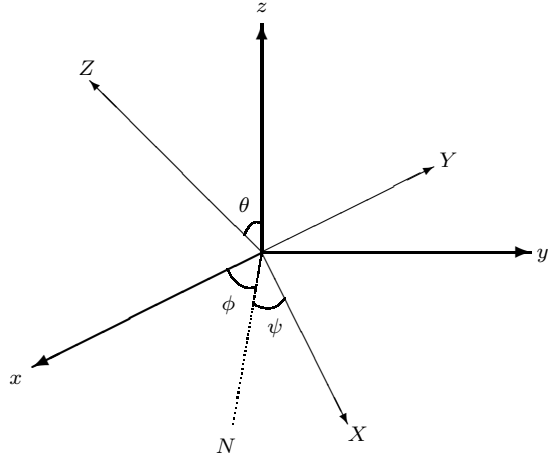
\begin{figure}
\begin{center}
\begin{picture}(100,100)

\thicklines
\put(50,50){\vector(-2,-1){40}}
\put(6,27){$x$}
\put(50,50){\vector(1,0){47}}
\put(98,49){$y$}
\put(50,50){\vector(0,1){40}}
\put(49,92){$z$}

\thinlines
\put(50,50){\vector(1,-2){15}}
\put(65,17){$X$}
\put(50,50){\vector(2,1){30}}
\put(81,65){$Y$}
\put(50,50){\vector(-1,1){30}}
\put(18,81){$Z$}

\put(42,15){$N$}
\qbezier[50](50,50)(49,44)(45,20)

\qbezier(50, 55)( 48,56)(47,53)
\put(46,57){$\theta$}
\qbezier(44,47)(46,43)(49,44)
\put(43,40){$\phi$}
\qbezier(48.5,41)(52,39)(54,42)
\put(51,36){$\psi$}

\end{picture}
\end{center}

\caption{The orientation of our body is specified by the three standard Euler angles $(\theta, \phi, \psi)$, as labelled above.  The fixed inertial-frame axes are denoted by $(x,y,z)$, while the rotating body-frame axes are denoted by $(X,Y,Z)$.  The so-called line of nodes, $N$, lies along the intersection of the $xy$ and $XY$ planes. }
\label{fig:euler_angles}
\end{figure}

We will begin by following \citet{shah77}, extending his treatment
slightly by modelling a star consisting of a triaxial rigid `crust'
and a pinned superfluid component of spherical moment of inertia
(Shaham assumed a biaxial crust).  Note that here, and for the entirety of this paper, by `crust' we mean not just
the outer solid phase but all parts of the star apart from the pinned
superfluid.  Referring our equations to the crust's body frame and
denoting by $\{n^1_a, n^2_a, n^3_a\}$ unit vectors along the crust's
principal axes, the crust's moment of inertia tensor is
\begin{equation}
\label{eq:crust_moi}
I^{\rm C}_{ab} = I^{\rm C}_1 n^1_a n^1_b +  I^{\rm C}_2 n^2_a n^2_b +
 I^{\rm C}_3 n^3_a n^3_b ,
\end{equation}
while the superfluid's is
\begin{equation}
\label{eq:SF_moi}
I^{\rm SF}_{ab} = I^{\rm SF} \delta_{ab} .
\end{equation}
The star's total angular momentum is then the sum of the angular
momenta of the two components:
\begin{equation}
\label{eq:J_total}
J_a = I^{\rm C}_{ab} \Omega^{\rm C}_b + I^{\rm SF} \Omega^{\rm SF}_a ,
\end{equation}
where $\Omega^{\rm C}_a$ and $\Omega^{\rm SF}_a$ are the crustal and
superfluid angular velocities, respectively.  Writing out the
components of $J_a$ explicitly:
\begin{eqnarray}
\label{eq:J1_rigid}
J_1 &=& I^{\rm C}_1 \Omega^{\rm C}_1 + I^{\rm SF} \Omega^{\rm SF}_1 , \\
\label{eq:J2_rigid}
J_2 &=& I^{\rm C}_2 \Omega^{\rm C}_2 + I^{\rm SF} \Omega^{\rm SF}_2 , \\
\label{eq:J3_rigid}
J_3 &=& I^{\rm C}_3 \Omega^{\rm C}_3 + I^{\rm SF} \Omega^{\rm SF}_3 .
\end{eqnarray}
If the pinned superfluid has a rotation axis given by the spherical polar angles
$(\theta^{\rm SF}, \phi^{\rm SF})$ with respect to the crust frame then
\begin{equation}
\label{eq:SF_components}
\Omega^{\rm SF}_a = \Omega^{\rm SF} 
( \sin\theta^{\rm SF}\cos\phi^{\rm SF} n^1_a
+ \sin\theta^{\rm SF}\sin\phi^{\rm SF} n^2_a
+ \cos\theta^{\rm SF} n^3_a) .
\end{equation}
The Euler equation of motion is
\begin{equation}
\label{eq:Euler}
\frac{d J_a}{dt} + \epsilon_{abc} \Omega^{\rm C}_b J_c = 0,
\end{equation}
where the time derivative is evaluated in the crust frame.  Writing out the
components explicitly we obtain:
\begin{eqnarray}
\label{eq:Euler_component1}
I^{\rm C}_1 \dot\Omega^{\rm C}_1 
+ \Omega^{\rm C}_2 \Omega^{\rm C}_3 (I^{\rm C}_3-I^{\rm C}_2)
+(\Omega^{\rm C}_2 \cos\theta^{\rm SF} 
  - \Omega^{\rm, C}_3 \sin\theta^{\rm SF}\sin\phi^{\rm SF}
)I^{\rm SF}\Omega^{\rm SF} &=& 0 , \\
\label{eq:Euler_component2}
I^{\rm C}_2 \dot\Omega^{\rm C}_2 
+ \Omega^{\rm C}_3 \Omega^{\rm C}_1 (I^{\rm C}_1-I^{\rm C}_3)
+(\Omega^{\rm C}_3 \sin\theta^{\rm SF}\cos\phi^{\rm SF} 
  - \Omega^{\rm C}_1 \cos\theta^{\rm SF}
)I^{\rm SF}\Omega^{\rm SF} &=& 0 , \\
\label{eq:Euler_component3}
I^{\rm C}_3 \dot\Omega^{\rm C}_3 
+ \Omega^{\rm C}_1 \Omega^{\rm C}_2 (I^{\rm C}_2-I^{\rm C}_1)
+(\Omega^{\rm C}_1 \sin\theta^{\rm SF}\sin\phi^{\rm SF} 
  - \Omega^{\rm, C}_2 \sin\theta^{\rm SF}\cos\phi^{\rm SF}
)I^{\rm SF}\Omega^{\rm SF} &=& 0 .
\end{eqnarray}
The non-precessing solution is obtained by setting $dJ_a/dt = 0$ in
equation (\ref{eq:Euler}), which in turn implies $\epsilon_{abc}
  \Omega^{\rm C}_b J_c = 0$, i.e. $\Omega^{\rm C}_a$ is parallel to
  $J_a$, so that the crust rotates steadily about the fixed angular
  momentum vector.  Equations
  (\ref{eq:Euler_component1})--(\ref{eq:Euler_component3}) then reduce
  to
\begin{eqnarray}
\label{eq:non_prec_component1}
\Omega^{\rm C}_2 \Omega^{\rm C}_3 (I^{\rm C}_3-I^{\rm C}_2)
+(\Omega^{\rm C}_2 \cos\theta^{\rm SF} 
  - \Omega^{\rm, C}_3 \sin\theta^{\rm SF}\sin\phi^{\rm SF}
)I^{\rm SF}\Omega^{\rm SF} &=& 0 , \\
\label{eq:non_prec_component2}
\Omega^{\rm C}_3 \Omega^{\rm C}_1 (I^{\rm C}_1-I^{\rm C}_3)
+(\Omega^{\rm C}_3 \sin\theta^{\rm SF}\cos\phi^{\rm SF} 
  - \Omega^{\rm C}_1 \cos\theta^{\rm SF}
)I^{\rm SF}\Omega^{\rm SF} &=& 0 , \\
\label{eq:non_prec_component3}
\Omega^{\rm C}_1 \Omega^{\rm C}_2 (I^{\rm C}_2-I^{\rm C}_1)
+(\Omega^{\rm C}_1 \sin\theta^{\rm SF}\sin\phi^{\rm SF} 
  - \Omega^{\rm, C}_2 \sin\theta^{\rm SF}\cos\phi^{\rm SF}
)I^{\rm SF}\Omega^{\rm SF} &=& 0 .
\end{eqnarray}
If we regard $\Omega^{\rm SF}, \theta^{\rm SF}, \phi^{\rm SF}$ as
fixed, as is appropriate for a perfectly pinned superfluid, these
equations then have a one-parameter family of solutions, which we will
parameterise in terms of $\Omega^{\rm C}_3$.  Then we find
\begin{eqnarray}
\label{eq:Omega1}
\Omega^{\rm C}_1 &=& \Omega^{\rm C}_3 
\frac{I^{\rm SF}\Omega^{\rm SF} \sin\theta^{\rm SF} \cos\phi^{\rm SF}}
{I_{31} \Omega^{\rm C}_3 + I^{\rm SF}\Omega^{\rm SF} \cos\theta^{\rm SF}} ,\\
\label{eq:Omega2}
\Omega^{\rm C}_2 &=& \Omega^{\rm C}_3 
\frac{I^{\rm SF}\Omega^{\rm SF} \sin\theta^{\rm SF} \sin\phi^{\rm SF}}
{I_{32} \Omega^{\rm C}_3 + I^{\rm SF}\Omega^{\rm SF} \cos\theta^{\rm SF}} ,
\end{eqnarray}
having defined
\begin{eqnarray}
\label{eq:I_31}
I_{31} &\equiv& I^{\rm C}_3 - I^{\rm C}_1 , \\
\label{eq:I_32}
I_{32} &\equiv& I^{\rm C}_3 - I^{\rm C}_2 , \\
\label{eq:I_21}
I_{21} &\equiv& I^{\rm C}_2 - I^{\rm C}_1 .
\end{eqnarray}
To fix the orientation of the body with respect to the inertial frame,
define a set of three Euler angles $(\theta, \phi, \psi)$ which
connect the body frame to the inertial frame in the standard manner,
as described in \citet{ll76} and illustrated in Figure \ref{fig:euler_angles}. 
Then
\begin{eqnarray}
\label{eq:Omega1_definition}
\Omega^{\rm C}_1 &=& \dot\phi \sin\theta \sin\psi + \dot\theta \cos\psi, \\
\label{eq:Omega2_definition}
\Omega^{\rm C}_2 &=& \dot\phi \cos\theta \cos\psi - \dot\theta \sin\psi, \\
\label{eq:Omega3_definition}
\Omega^{\rm C}_3 &=& \dot\phi \cos\theta + \dot\psi .
\end{eqnarray}
Our non-precessing solution has all components of $\Omega^{\rm C}_a$
constant, and so equations (\ref{eq:Omega1_definition}) and
(\ref{eq:Omega2_definition}) imply that $\theta$ and $\psi$ must be
constant, leaving
\begin{eqnarray}
\Omega^{\rm C}_1 &=& \dot\phi \sin\theta \sin\psi, \\
\Omega^{\rm C}_2 &=& \dot\phi \cos\theta \cos\psi, \\
\Omega^{\rm C}_3 &=& \dot\phi \cos\theta.
\end{eqnarray}
These can be inverted easily to give
\begin{eqnarray}
\label{eq:tan_theta_def}
\tan\theta &=& 
\frac{\left[(\Omega^{\rm C}_1)^2 + (\Omega^{\rm C}_2)^2\right]^{1/2}}
{\Omega^{\rm C}_3} , \\
\label{eq:phi_def}
\phi &=& \Omega^{\rm C}t + \phi_0 , \\
\label{eq:tan_phi_def}
\tan\psi &=& \frac{\Omega^{\rm C}_1}{\Omega^{\rm C}_2} .
\end{eqnarray}
The angles $\theta$ and $\psi$ are constant, while $\phi$ is a
linearly increasing function of time---it is this angle that generates
the rotation.  Elimination of $\Omega^{\rm C}_1$ and $\Omega^{\rm
  C}_2$ in favour of $\Omega^{\rm C}_3$ using equations
(\ref{eq:Omega1}) and (\ref{eq:Omega2}) gives
\begin{eqnarray}
\label{eq:tan_theta}
\tan\theta &=& \tan\theta^{\rm SF} 
\left[
\cos^2\phi^{\rm SF}
\left(1+\frac{I_{31}\Omega^{\rm C}_3}{I^{\rm SF}\Omega^{\rm SF}_3}
\right)^{-2}
+
\sin^2\phi^{\rm SF}
\left(1+\frac{I_{32}\Omega^{\rm C}_3}{I^{\rm SF}\Omega^{\rm SF}_3}
\right)^{-2}
\right]^{1/2} , \\
\label{eq:phi}
\phi &=& \Omega^{\rm C}_3 t 
\left\{
\tan^2\theta^{\rm SF}
\left[
\cos^2\phi^{\rm SF}
\left(1+\frac{I_{31}\Omega^{\rm C}_3}{I^{\rm SF}\Omega^{\rm SF}_3}
\right)^{-2}
+
\sin^2\phi^{\rm SF}
\left(1+\frac{I_{32}\Omega^{\rm C}_3}{I^{\rm SF}\Omega^{\rm SF}_3}
\right)^{-2} 
\right]
+ 1
\right\}^{1/2} + \phi_0, \\
\label{eq:tan_psi}
\tan\psi &=& \tan(\pi/2-\phi^{\rm SF})
\left[
1+ \frac{I_{32}\Omega^{\rm C}_3}{I^{\rm SF}\Omega^{\rm SF}_3}
\right]
\left[
1+ \frac{I_{31}\Omega^{\rm C}_3}{I^{\rm SF}\Omega^{\rm SF}_3}
\right]^{-1} ,
\end{eqnarray}
where $\phi_0$ is a constant.  Clearly, there is a rather complicated
relationship between the orientation of the body with respect to the
inertial frame and the parameters $\Omega^{\rm C}_3, \Omega^{\rm SF},
\theta^{\rm SF}, \phi^{\rm SF}$.  

However, we can simplify the above.  The asymmetries $I_{31}, I_{32}$
are sourced by crustal strains or magnetic fields, and so are likely
to be very small.  In the case of crustal strains, \citet{hetal06},
following the work of \citet{ucb00}, estimated \emph{maximum}
deformations of approximately
\begin{equation}
\frac{\Delta I}{10^{45} \, \rm g \, cm^2} \approx 3 \times 10^{-5} 
\left(\frac{u_{\rm break}}{0.1}\right) ,
\end{equation}
where $u_{\rm break}$ denotes the breaking strain of the crust.  Recent molecular dynamics calculations \citep{hk09} estimate this breaking to strain to be approximately  $0.1$.  Note that the above crustal deformation is an upper bound as it assumes the crust is maximally strained.  In the case of an internal magnetic field of strength $\sim B$ a naive estimate is
\citep{jone02}
\begin{equation}
\frac{\Delta I}{10^{45} \, \rm g \, cm^2} \approx 10^{-12} 
\left(\frac{B}{10^{12} \, \rm G}\right)^2 ,
\end{equation}
although more detailed calculations show the result depends on the
assumed field geometry \citep{bg96, hetal08, cfgp09, lj09} and on the
superconducting nature of the core (see \citet{cutl02} and references
therein).
 
On the other hand, the moment of inertia of the superfluid is
relatively large: $I_{\rm SF}/I \sim 10^{-2}$ for crustal pinning,
$I^{\rm SF}/I \sim 1$ for core pinning \citep{saul89}.  We can
therefore make use of the inequalities
\begin{equation}
\label{eq:small_deformations}
\frac{I_{31} \Omega^{\rm C}_3}{I^{\rm SF}\Omega^{\rm SF}_3} \ll 1
\hspace{10mm}
\frac{I_{32} \Omega^{\rm C}_3}{I^{\rm SF}\Omega^{\rm SF}_3} \ll 1
\end{equation}
to simplify our expressions for the Euler angles
\begin{eqnarray}
\label{eq:approx_theta}
\theta &\approx& \theta^{\rm SF} , \\
\label{eq:approx_phi}
\phi &\approx& \frac{\Omega^{\rm C}_3 t}{\cos\theta^{\rm SF}} + \phi_0, \\
\label{eq:approx_psi}
\psi &\approx& \pi/2 - \phi^{\rm SF} ,
\end{eqnarray}
This shows that the star rotates about an axis very close to $n^{\rm
  SF}_a$.  So, after having gone through a careful analysis, the basic
picture is very simple: the star rotates steadily about an axis very
close to the superfluid pinning axis.  From this point on we will
simplify our notation slightly, writing $(\Omega^{\rm C}_a \Omega^{\rm
  C}_a)^{1/2}$ simply as $\Omega$, the observed stellar spin
frequency.

\section{Gravitational wave emission}
\label{sect:gwe}

\subsection{Calculation of the gravitational wave signal}

Having obtained the Euler angles $(\theta, \phi, \psi)$ giving the
orientation of the body with respect to the inertial frame it is
straightforward to calculate the corresponding gravitational wave
emission.  To do so we will calculate the source's multipole moments,
as described by \citet{thor80}; this will make clear the separation
between the frequency harmonics and will also be useful in discussing
maximum possible wave amplitudes in section \ref{sect:gwa}.  The
multipole moments of the source are complex scalars defined by
integrals over the star's density field $\rho$:
\begin{equation}
Q_{lm} = \int \rho r^l Y^\ast_{lm} \, d^3 x .
\end{equation}
When the density field is written as the sum $\rho = \Sigma \delta
\rho_{lm} Y_{lm}$ this reduces to
\begin{equation}
Q_{lm} = \int \delta \rho_{lm}(r) r^{l+2} \, dr .
\end{equation}
When the body has orientation $\theta, \phi, \psi$ relative to the
inertial axes we find
\begin{eqnarray}
Q_{21} &=& \frac{1}{2} \sqrt{\frac{15}{2\pi}} e^{-i\phi} 
[-I_{21} \sin2\psi \sin\theta - i \sin2\theta (I_{21}\cos^2\psi - I_{31})] , \\
Q_{22} &=& \frac{1}{2} \sqrt{\frac{15}{2\pi}} e^{-2i\phi} 
[I_{21}(\cos^2\psi \cos^2\theta - \sin^2\psi) + I_{31}\sin^2\theta
-i I_{21}\sin2\psi \cos\theta] .
\end{eqnarray}
The $Q_{21}$ multipole moment generates radiation at the spin
frequency $\Omega$, while the $Q_{22}$ multipole moment generates
radiation at $2\Omega$. 

In calculating the gravitational wave emission, we will consider an
observer a distance $r$ from the star whose coordinate system
$(X,Y,Z)$ is obtained from the inertial $(x,y,z)$ frame by a rotation
through an inclination angle $\iota$ about the inertial $x$-axis,
followed by a translation through a distance $r$ along $OZ$.  The
gravitational wave then propagates along $OZ$ and, within the mass
quadrupole approximation to general relativity, the two polarisations
which make up the transverse traceless wave field are
\begin{eqnarray}
\label{eq:h_plus_omega}
h_+^\Omega &=& -\frac{\Omega^2}{r} \sin\iota \cos\iota
\big\{[I_{21}\cos^2\psi - I_{31}]\sin2\theta \cos\phi
- I_{21} \sin 2\psi \sin\theta \sin\phi \big\} , \\
\label{eq:h_cross_omega}
h_\times^\Omega &=& -\frac{\Omega^2}{r} \sin\iota
\big\{I_{21} \sin 2\psi \sin\theta \cos\phi 
+ (I_{21}\cos^2\psi - I_{31}) \sin 2\theta \sin\phi \big\} , \\
\label{eq:h_plus_2omega}
h_+^{2\Omega} &=& -\frac{2\Omega^2}{r} (1+\cos^2\iota)
\big\{[I_{21}(\cos^2\psi \cos^2\theta - \sin^2\psi) 
+ I_{31}\sin^2\theta]\cos2\phi
- I_{21} \sin 2\psi \cos\theta \sin2\phi \big\} , \\
\label{eq:h_cross_2omega}
h_\times^{2\Omega} &=& -\frac{2\Omega^2}{r} 2\cos\iota
\big\{I_{21} \sin 2\psi \cos\theta \cos2\phi 
+  [(I_{21}(\cos^2\psi \cos^2\theta - \sin^2\psi) 
+ I_{31} \sin^2\theta ] \sin 2\phi \big\}.
\end{eqnarray}
These equations, together with the equations
(\ref{eq:tan_theta})--(\ref{eq:tan_psi}) (or alternatively the
approximate equations (\ref{eq:approx_theta})--(\ref{eq:approx_psi}))
then give a complete specification of the gravitational field.  As
anticipated, there are harmonics at both $\Omega$ and $2\Omega$.

\subsection{Counting the number of parameters}

The actual signal received by a detector can be written as
\begin{equation}
\label{eq:h_F}
h(t) = F_+(t) h_+(t) + F_\times(t)h_\times(t) ,
\end{equation}
where $F_+$ and $F_\times$ are the beam pattern factors whose form can
be found in \citet{jks98}.  They depend upon the detector location,
source position on the sky, and on the orientation of the source's
spin axis.  This introduces three unknown parameters over and above
those appearing in the expressions for $h_+$ and $h_\times$,
specifically the source's location on the sky (which can be specified
by right ascension $\alpha$ and declination $\delta$), and an angle
giving the projection of the source's spin axis on the plane
perpendicular to the line-of-sight (which can be specified by the
polarisation angle $\psi_{\rm pol}$).

We can now count the number of parameters required to reconstruct the
full signal.  Let us first examine two special cases, to demonstrate
consistency with previously published results.

If we put $\theta=0$ in
(\ref{eq:h_plus_omega})--(\ref{eq:h_cross_2omega}) we obtain
\begin{eqnarray}
\label{eq:h_plus_triaxial}
h_+ &=& -\frac{2\Omega^2}{r} I_{21} (1+\cos^2 i) \cos 2(\phi+\psi) , \\
\label{eq:h_cross_triaxial}
h_\times &=& -\frac{2\Omega^2}{r} I_{21} 2\cos i \sin 2(\phi+\psi) .
\end{eqnarray}
This corresponds to a triaxial star spinning about a principal axis;
it is the sort of signal that has been assumed in many gravitational wave
searches (see e.g. \citet{aetal07_PRD76a}).  A possible choice of
parameters to describe the signal is:
\begin{equation}
  \left\{\frac{\Omega^2 I_{21}}{r}, \Omega, 
    (\phi+\psi)_0, \iota, \psi_{\rm pol}, \alpha, \delta \right\} .
\end{equation}
In this case there are 7 parameters (note that in this case the angles
$\phi$ and $\psi$ are degenerate, and only their sum appears in the
waveform).  There is radiation only at the $2\Omega$ harmonic.

If we put $I_{21}=0$ in
(\ref{eq:h_plus_omega})--(\ref{eq:h_cross_2omega}) we obtain
\begin{eqnarray}
\label{eq:h_plus_biaxial}
h_+^\Omega &=& \frac{\Omega^2}{r} \sin\iota \cos\iota I_{31} 
\sin 2\theta \cos\phi ,\\
h_\times^\Omega &=& \frac{\Omega^2}{r} \sin\iota I_{31} 
\sin 2\theta \sin\phi ,\\
h_+^{2\Omega} &=& -\frac{2\Omega^2}{r} (1+\cos^2\iota) I_{31} 
\sin^2\theta \cos2\phi ,\\
\label{eq:h_cross_biaxial}
h_\times^{2\Omega} &=& -\frac{2\Omega^2}{r} 2\cos\iota I_{31} 
\sin^2\theta \sin2\phi .
\end{eqnarray}
This is identical to the wave field of a precessing biaxial star, and
therefore is of exactly the same form as given in \citet{zs79} and
\citet{jks98}.  Note, however, that in our case the motion is not one
of free precession---the star rotates steadily about the fixed angular
momentum vector, with no superimposed rotation about the crust's
symmetry axis.  A possible choice of parameters is:
\begin{equation}
  \left\{\frac{\Omega^2 I_{31}}{r}, \Omega, 
    \phi_0, \theta, \iota, \psi_{\rm pol}, \alpha, \delta \right\} .
\end{equation}
In this case there are 8 parameters, with the angle $\theta$ being the
extra parameter as compared to the case considered above.  There is
radiation at both the $\Omega$ and $2\Omega$ harmonics, and $\theta$
controls the relative strength of these.  In the limit of small
$\theta$, they scale as $\theta$ and $\theta^2$ respectively, so the
$\Omega$ harmonic can dominate, but both go to zero (at fixed
$I_{31}$) in the limit $\theta \rightarrow 0$.

Finally, consider the general case.  The wave field is then as given
in equations (\ref{eq:h_plus_omega})--(\ref{eq:h_cross_2omega}).  A
possible choice of parameters is
\begin{equation}
\label{eq:10_params}
  \left\{\frac{\Omega^2 I_{21}}{r}, \frac{I_{31}}{I_{21}}, 
    \Omega, \phi_0, \theta, \psi, \iota, \psi_{\rm pol}, 
    \alpha, \delta \right\} .
\end{equation}
In this case there are 10 parameters, with the ratio $I_{31}/I_{21}$
and the angle $\psi$ being the extra parameters as compared to the
case above.  In general, there is radiation at both $\Omega$ and
$2\Omega$, but the relative strengths depends on the parameters in a
rather complicated way.  Let us consider the extremes.  There are
solutions where the star radiates only at $2\Omega$, with the signal
being proportional to one of $I_{21}, I_{31}, I_{32}$.  These
correspond to the familiar case of rotation of a triaxial star about a
principal axis.  More interestingly, there also exist solutions where
the star radiates only at $\Omega$.  The solutions are:
\begin{eqnarray}
  \sin\psi=0, \hspace{5mm} \tan\theta = \sqrt{-\frac{I_{21}}{I_{31}}}, &&
  |Q_{21}| = \sqrt{\frac{15}{2\pi}} \sqrt{-I_{21}I_{31}} ,\\
  \cos\psi=0, \hspace{5mm} \sin\theta = \sqrt{+\frac{I_{21}}{I_{31}}}, &&
  |Q_{21}| = \sqrt{\frac{15}{2\pi}} \sqrt{I_{21}I_{32}} ,\\
  \cos\theta=0, \hspace{5mm} \sin\psi = \sqrt{+\frac{I_{31}}{I_{21}}}, &&
  |Q_{21}| = \sqrt{\frac{15}{2\pi}} \sqrt{I_{31}I_{32}} .
\end{eqnarray}
Clearly, to obtain radiation only at the $\Omega$ harmonic, special values of the parameters are necessary.  For general values of
the parameters, there will be radiation at both harmonics.

\subsection{Gravitational wave amplitudes}
\label{sect:gwa}

Let us now examine the relative strengths at which the two quadrupole
moments can radiate.  As noted by \citet{ucb00}, the $Q_{22}$
perturbations are somewhat more efficient gravitational wave emitters,
in the sense that for a given magnitude of the quadrupole moment (i.e.
$|Q_{21}| = |Q_{22}|$), the energy flux in the $2\Omega$ radiation is
greater than that in the $\Omega$ radiation.  To see this we can
calculate the inclination-angle ($\iota$) dependent gravitational wave
energy flux using \citep{thor80}
\begin{equation}
\label{eq:flux_formula}
F = \frac{1}{32\pi} < \dot h_{ab}^{\rm TT} \dot h_{ab}^{\rm TT} > ,
\end{equation}
where the angle brackets denote a time average over several periods.
This gives the following when written in terms of multipole moments:
\begin{eqnarray}
F_\Omega (\iota) &=& \frac{1}{60} \frac{\Omega^6}{r^2} |Q_{21}|^2 
\sin^2\iota (1+\cos^2\iota) , \\
F_{2\Omega} (\iota)  &=& \frac{4}{15} \frac{\Omega^6}{r^2} |Q_{22}|^2 
[(1+\cos^2\iota)^2 + 4\cos^2\iota] .
\end{eqnarray}
When integrated over the sphere the corresponding total luminosities
are
\begin{eqnarray}
\label{eq:E_dot_Omega}
\dot E_\Omega &=& \frac{4\pi}{75} \Omega^6 |Q_{21}|^2 , \\
\label{eq:E_dot_2Omega}
\dot E_{2\Omega} &=& \frac{256\pi}{75} \Omega^6 |Q_{22}|^2 . 
\end{eqnarray}
As noted by \citet{ucb00}, for a given magnitude of the quadrupole
moment, a $Q_{22}$ perturbation is $64$ times more luminous than a
$Q_{21}$ perturbation.

However, in terms of detectability, the relevant quantity is not the
energy flux but the signal-to-noise ratio, which is linear in the
gravitational wave amplitude \citep{jks98}.  For a star radiating
continuously at gravitational wave frequencies $\Omega$ and $2\Omega$,
\citet{jks98} showed that the signal-to-noise is
\begin{equation}
d = \sqrt{d_1^2 + d_2^2} ,
\end{equation}
where 
\begin{eqnarray}
d_1 &=& \frac{2}{S_h(\Omega)} \int h_1^2(t) \, dt ,\\
d_2 &=& \frac{2}{S_h(2\Omega)} \int h_2^2(t) \, dt .
\end{eqnarray}
In this equation, the time integral is over the duration of the
observation, $S_h(f)$ is the detector's noise spectral density, and
$h_1(t), h_2(t)$ are defined in terms of the beam functions of
equation (\ref{eq:h_F}):
\begin{eqnarray}
h_1(t) &=& F_+(t) h_+^\Omega(t) + F_\times(t) h_\times^\Omega(t) \\
h_2(t) &=& F_+(t) h_+^{2\Omega}(t) + F_\times(t) h_\times^{2\Omega}(t) .
\end{eqnarray}
The wave amplitudes $h_+^\Omega(t), h_\times^\Omega(t),
h_+^{2\Omega}(t), h_\times^{2\Omega}(t)$ are given in equations
(\ref{eq:h_plus_omega})--(\ref{eq:h_cross_2omega}).  In the case of
gravitational wave searches at a single frequency from a known pulsar,
it was found that a signal-to-noise threshold of $11.4$ was necessary
to achieve a false alarm probability of $1\%$ and a false dismissal
probability of $10\%$ \citep{aetal04_PRD69}.  This signal-to-noise
threshold will have to be recalculated for a two-frequency search; we
will discuss this further in section \ref{sect:discussion}.

\begin{figure}
\centerline{\includegraphics[height= 6.5cm,clip]{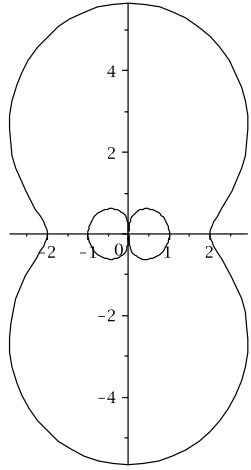}}
\caption{Contour plot showing normalised signal strengths of the
  $\Omega$ (inner contour) and $2\Omega$ (outer contour) harmonics,
  assuming $|Q_{21}| = |Q_{22}|$.  For the $\Omega$ radiation we plot
  $\sin \iota (1+\cos^\iota)^{1/2}$, for the $2\Omega$ radiation
  $2[(1+\cos^\iota)^2 + 4\cos^2\iota]^{1/2}$.  The rotation axis
  ($\iota = 0$) lies along the vertical.}
\label{fig:polar_detectability}
\end{figure}

The value of the signal-to-noise for a particular star, and the
relative importance of the two harmonics, depends upon all 10 of the
parameters given in equation (\ref{eq:10_params}).  In order to gain
a \emph{rough} idea of the relative importance of the two harmonics,
we will define an approximate signal-to-noise ratio for each
according to
\begin{equation}
  \rho = \frac{h\sqrt{T_{\rm obs}}}{\sqrt{S_h (f_{\rm GW})}} ,
\end{equation}
where $T_{\rm obs}$ is the duration of the observation (see
\citet{jks98} for a justification of the scaling with observation
time).  For the gravitational wave amplitude $h$ we will take the
square root of the time-averaged sum of the squares of the plus and
cross gravitational wave polarisations, i.e.
\begin{equation}
h = <(h_+)^2 + (h_\times)^2>^{1/2} ,
\end{equation}
where the angle brackets denote time averaging.  Then
equation (\ref{eq:flux_formula}) gives
\begin{equation}
\label{eq:defn_h}
h \equiv <(h_+)^2 + (h_\times)^2>^{1/2} = \frac{4}{\Omega_{\rm GW}} \sqrt{\pi F} ,
\end{equation}
allowing us to write the signal-to-noise ratios for the two harmonics
as
\begin{eqnarray}
\label{eq:rho_Omega}
\rho_\Omega (\iota) &=& \frac{A}{\sqrt{S_h(\Omega)}} |Q_{21}| \,
\sin \iota (1+\cos \iota)^{1/2} , \\
\label{eq:rho_2Omega}
\rho_{2\Omega} (\iota) &=& \frac{2A}{\sqrt{S_h(2\Omega)}} |Q_{22}| \, 
[(1+\cos \iota)^2 + 4\cos^2\iota]^{1/2} ,
\end{eqnarray}
where
\begin{equation}
A = 2 \left(\frac{\pi}{15}\right)^{1/2} T_{\rm obs}^{1/2} \Omega^2 .
\end{equation}
To compare the relative signal strengths of these harmonics, the
inclination angle-dependent coefficients of $A |Q|^2/S_h(f)$ appearing
in the above equations are presented in Figure
\ref{fig:polar_detectability}.
The $Q_{21}$ multipole emits preferentially in the rotational equator
($\iota = \pi/2$), emitting no radiation along the rotation axis
($\iota = 0$), while the $Q_{22}$ emits preferentially along the
rotation axis, with a non-zero wave amplitude in all directions.
Clearly, for all observer locations and for $|Q_{21}| = |Q_{22}|$, the
$2\Omega$ radiation is stronger than the $\Omega$ radiation.  The
ratio of the sky-averaged gravitational waves strengths (as defined in
equation (\ref{eq:defn_h})) is approximately $3.9 \approx 4$.  Of
course, the ratio of the possible maximum signal-to-noise ratios would
also fold in the frequency dependence of the detector noise between
frequencies $\Omega$ and $2\Omega$.

If instead we were to place an upper limit on the wave amplitude of a
star known to be losing kinetic energy at a rate $\dot E$, then
equations (\ref{eq:E_dot_Omega}) and (\ref{eq:E_dot_2Omega}) show that
the upper limits on the multipoles stand in the ratio $|Q_{21}^{\rm
  max}| = 8 |Q_{22}^{\rm max}|$ \citep{ucb00}.  It then follows that a
star losing energy at a known rate is, when sky-averaged, $8/3.9
\approx 2$ times stronger in gravitational wave amplitude if it
emitting via a pure $Q_{21}$ multipole than when emitting via a
$Q_{22}$ multipole.  Such an upper limit is appropriate for known
pulsars of measured spin-down rate (if one assumes all of the kinetic energy being lost is converted into gravitational wave emission),    or for low-mass X-ray binaries of known
X-ray luminosity (if one assumes the spin-up accretion torque, as estimated from the X-ray luminosity,  is  balanced by a spin-down gravitational torque).

\section{Extension to an elastic star}
\label{sect:etaes}

It is straightforward to allow for the rotationally-induced
deformation of the star by adding terms to the moment of inertia
tensor.  For the crust we amend equation (\ref{eq:crust_moi}) to
\begin{equation}
\label{eq:elastic_crust_moi}
I^{\rm C}_{ab} = I^{\rm C}_1 n^1_a n^1_b +  I^{\rm C}_2 n^2_a n^2_b +
 I^{\rm C}_3 n^3_a n^3_b + \Delta I^{\rm C}_\Omega n^{\rm C}_a n^{\rm C}_b ,
\end{equation}
where $n^{\rm C}_a$ is a unit vector pointing along $\Omega^{\rm
  C}_a$.  Physically, the term $\Delta I^{\rm C}_\Omega n^{\rm C}_a
n^{\rm C}_b$ represents a deformation biaxial about $\Omega^{\rm
  C}_a$, and $\Delta I^{\rm C}_\Omega$ is numerically equal to the
difference in moments of inertia about axes parallel and orthogonal to
the $\Omega^{\rm C}_a$ induced by the rotation.  The terms in $I^{\rm
  C}_1$, $I^{\rm C}_2$ and $I^{\rm C}_3$ give the moment of inertia of
the star in the absence of rotation.  The differences $I_{31},
I_{32}, I_{21}$ are defined as before (equations
(\ref{eq:I_31})--(\ref{eq:I_21})) and can be non-zero only by virtue
of non-rotational forces, e.g. elastic or magnetic strains.

A similarly decomposition applies for the superfluid:
\begin{equation}
\label{eq:fluid_SF_moi}
I^{\rm SF}_{ab} = I^{\rm SF} \delta_{ab} 
+ \Delta I^{\rm SF}_\Omega n^{\rm SF}_a n^{\rm SF}_b ,
\end{equation}
where $n^{\rm SF}_a$ is a unit vector along $\Omega^{\rm SF}_a$.  The
star's angular momentum components then become
\begin{eqnarray}
\label{eq:J1_elastic}
J_1 &=& (I^{\rm C}_1+\Delta I^{\rm C}_\Omega) \Omega^{\rm C}_1 
+ (I^{\rm SF} + \Delta I^{\rm SF}_\Omega) \Omega^{\rm SF}_1 , \\
\label{eq:J2_elastic}
J_2 &=& (I^{\rm C}_2+\Delta I^{\rm C}_\Omega) \Omega^{\rm C}_2 
+ (I^{\rm SF} + \Delta I^{\rm SF}_\Omega) \Omega^{\rm SF}_2 , \\
\label{eq:J3_elastic}
J_3 &=& (I^{\rm C}_3+\Delta I^{\rm C}_\Omega) \Omega^{\rm C}_3 
+ (I^{\rm SF} + \Delta I^{\rm SF}_\Omega) \Omega^{\rm SF}_3 .
\end{eqnarray}
These are identical in form to the corresponding rigid star equations,
(\ref{eq:J1_rigid})--(\ref{eq:J3_rigid}), the mapping from the rigid
star to elastic star equations being $I^{\rm C}_1 \rightarrow I^{\rm
  C}_1 + \Delta I^{\rm C}_\Omega$ (and similarly for $I^{\rm C}_2,
I^{\rm C}_3$) and $I^{\rm SF} \rightarrow I^{\rm SF} + \Delta I^{\rm
  SF}_\Omega$.  It follows at once that the solutions, both
precessional and non-precessional, are of the same form for this more
realistic elastic model as for the rigid star.

There will however be a change in the corresponding gravitational wave
emission, as we must now include the contributions from the mass
asymmetries associated with the $\Delta I^{\rm C}_\Omega$ and $\Delta
I^{\rm SF}_\Omega$ components.  For the non-precessional solution it
is easy to convince oneself that the impact on the gravitational wave
emission is minimal.  The $\Omega^{\rm C}_a$ vector points along the
invariant angular momentum axis, so there is no gravitational wave
emission from the $\Delta I^{\rm C}_\Omega$ term.  The $\Delta I^{\rm
  SF}_\Omega$ piece represents a biaxial deformation about
$\Omega^{\rm SF}_a$, which is misaligned with the angular momentum
axis.  This piece therefore does radiate.  Rather than calculating
this contribution fully, we can make a simple estimate.  This
deformation is of size $\Delta I^{\rm SF}_\Omega$ and moves at a rate
$\Omega$ in a cone of half-angle $\cos^{-1}(n^J_a n^{\rm SF}_a)$ about
$n^J_a$, the unit vector in the direction of $J_a$.  Using the results
of section \ref{sect:bm}, a little algebra leads to
\begin{equation}
\cos^{-1}(n^J_a n^{\rm SF}_a) \approx 
\frac{\sin\theta^{\rm SF}
\big[\cos^2\theta^{\rm SF}(I_{31}\cos^2\phi^{\rm SF} + I_{32}\sin^2\phi^{\rm SF})^2
+(I_{21}\sin\theta^{\rm SF}\sin\phi^{\rm SF}\cos\phi^{\rm SF})^2\big]^{1/2}  
\Omega^{\rm C}_3} {I^{\rm SF} \Omega^{\rm SF}_3} ,
\end{equation}
where we have worked to leading order in the small quantities of
equation (\ref{eq:small_deformations}).  We see that $\Omega^{\rm
  SF}_a$ lies very close to $J_a$; for $\Omega^{\rm C}_3 \approx
\Omega^{\rm SF}_3$ we obtain
\begin{equation}
\label{eq:misalignment}
\cos^{-1}(n^J_a n^{\rm SF}_a) 
\sim \frac{\Delta I}{I^{\rm SF}} ,
\end{equation}
where we have defined
\begin{equation}
\Delta I = \max(I_{21}, I_{31}, I_{32}) .
\end{equation}
It follows that the $\Delta I^{\rm SF}_\Omega$ term produces a
gravitational wave field of a similar form to the $I_{31}$ term in
equations (\ref{eq:h_plus_biaxial})--(\ref{eq:h_cross_biaxial}), but
with the replacements $I_{31} \rightarrow \Delta I^{\rm SF}_\Omega$
and $\theta \rightarrow \Delta I/I^{\rm SF}$.  It follows that the
contribution of this component to the gravitational wave emission is
weaker than that of the crust as calculated in equations
(\ref{eq:h_plus_biaxial})--(\ref{eq:h_cross_biaxial}) by a factor
$\Delta I^{\rm SF}_\Omega/I^{\rm SF}$.  For all realistic scenarios
this ratio will be much less than unity, so we will therefore not
consider this component of the wave emission in the remainder of this
paper.  We therefore see that allowance for rotational deformation of
the non-precessing star has minimal impact on the gravitational wave
emission.

\section{Limitations and caveats} 
\label{sect:lac}

What limits the maximum gravitational wave amplitude?  In the absence
of the pinned superfluid component the answer has been obtained by
\citet{ucb00}: the $Q_{22}$ multipole is limited by the breaking
strain $u_{\rm break}$ of the crust:
\begin{equation}
Q_{\rm max} \approx 1.2 \times 10^{39} {\, \rm g \, cm^2 \,} 
\left(\frac{u_{\rm break}}{10^{-1}}\right) ,
\end{equation}
(see also \citet{hetal06}).  Also, as discussed in the conclusions
section of \citet{ucb00}, an identical limit is placed on $Q_{21}$.
(That these bounds are identical can be understood by noting that the
real parts of $Y_{22}$ and $Y_{21}$ density perturbations are
identical up to a rotation of axes).  So, if the ability of the
crustal strain to produce the asymmetries $I_{21}, I_{31}, I_{32}$
were the only factor, the upper bounds on the signal strengths would
be given by equation (\ref{eq:rho_Omega}) with $|Q_{21}| = Q_{\rm
  max}$ for a pure $Q_{21}$ deformation, and by equation
(\ref{eq:rho_2Omega}) with $|Q_{22}| = Q_{\rm max}$ for a pure
$Q_{22}$ deformation.  In the general case of a mixed $Q_{21}$,
$Q_{22}$ deformation, both perturbations contribute to the strain
field, and the limit on each is reduced \citep{ucb00}.

However, the presence of the pinned superfluid complicates this
picture.  In our model, there is an angular velocity misalignment
between the superfluid and the rest of the star, as calculated above
(equation \ref{eq:misalignment}).  It follows that over the pinning
region the crust exerts a force on the superfluid, supplying the
torque required to allow for its time-varying angular velocity.  As is
demonstrated in appendix \ref{sect:app_tct}, this torque is supplied
by the Magnus force, generated by the flow of superfluid past the
pinned vortices.  It follows that there are new failure mechanisms as
compared to the case studied by \citet{ucb00}.  Firstly, the pinning
force interaction between the solid crust and the vortices has to be
sufficiently strong that the vortices don't unpin.  Secondly, the
corresponding equal but opposite force on the crust will strain it
further.  This additional strain might push the crust closer to, or
beyond, its breaking strain.  This will also produce some change in the
moment of inertia tensor, potentially modifying the gravitational wave
emission.  Finally, there is also the possibility that relative flow
between the superfluid and the vortices may trigger an instability.
Let us make some simple estimates of the importance of these effects.

To do so, note that we could set our star up with an arbitrarily large
difference between $\Omega^{\rm SF}_a$ and $\Omega^{\rm C}_a$.
However, in the physically most plausible case, these two vectors will
be of equal magnitudes, but misaligned by the angle $\cos^{-1} (n^J_a
n^{\rm SF}_a)$, producing a vortex--superfluid relative velocity of
order
\begin{equation}
\Delta v \sim \Omega R \, \cos^{-1} (n^J_a n^{\rm SF}_a) ,
\end{equation}
where $R$ is the stellar radius.  Equation (\ref{eq:misalignment})
then leads to
\begin{equation}
  \label{eq:v_critical_nonparam}
  \Delta  v \sim \Omega R \frac{\Delta I}{I^{\rm SF}} .
\end{equation}
Parameterising:
\begin{equation}
  \label{eq:v_critical_param}
  \Delta v \sim 6 \times 10^4 {\, \rm cm \, s^{-1}} 
  \left(\frac{f}{\rm kHz}\right)
  \left(\frac{\Delta I/I}{10^{-7}}\right)
  \left(\frac{10^{-2}}{I^{\rm SF}/I}\right)
  \left(\frac{R}{10^6 \, \rm cm}\right) ,
\end{equation}
where $f=\Omega/(2\pi)$, and $I$ denotes the total stellar moment of
inertia.  This will be useful in interpreting some of the estimates
that follow.  This can equivalently be restated in terms of a pinning
force per unit length of vortex \citep{lc02}:
\begin{equation}
f_a = \rho^{\rm SF} \kappa \epsilon_{abc} n^{\rm SF}_b 
(v^{\rm SF}_c-v^{\rm C}_c)
\end{equation}
where $\rho^{\rm SF}$ is the superfluid density and $\kappa$ the
quantum of vorticity.  Writing the relative velocity using equation
(\ref{eq:v_critical_nonparam}) gives
\begin{equation}
  \label{eq:pinning_force}
  f \sim 1.3 \times 10^{16} {\, \rm dyn \, cm}^{-1} \,
  \left(\frac{f}{\rm kHz}\right)
  \left(\frac{\Delta I/I}{10^{-7}}\right)
  \left(\frac{10^{-2}}{I^{\rm SF}/I}\right)
  \left(\frac{\rho^{\rm SF}}{10^{14} \, \rm g \, cm^{-3}}\right)
  \left(\frac{R}{10^6 \, \rm cm}\right) .
\end{equation}
These equations give the relative vortex-superfluid velocities, or,
equivalently, the pinning forces per unit length of vortex, required
to sustain the non-precessional solutions described in this paper.

\subsection{Breaking the pinning}

First let us ask if the pinning is strong enough to withstand the
Magnus force.  There exist both observational and theoretical inputs
to this issue.  On the observational side, glitches and neutron star
precession may tell us something about the pinning strength, while, on
the theoretical side, there has been some microphysical modelling of
how vortices interact with the rest of the star.

Let us start with the observations of glitching pulsars. The smaller
glitches, such as those seen in the Crab, may be caused by
`starquakes', i.e.  changes in the elastic strain of the crust, and so
may not carry information about the superfluid pinning.  However, the
larger so-called `giant glitches', such as those in the Vela pulsar,
are believed to represent sudden unpinning events.  An estimate of the
relative superfluid-vortex velocity at unpinning comes from assuming
angular momentum conservation \citep{lc02, ga09}.  If the glitch
represents a sudden angular momentum exchange between the pinned
superfluid and the rest of the star (in our language the `crust'),
then the changes in angular frequency of these two components are
related via $I^{\rm SF} \Delta \Omega^{\rm SF} + I^{\rm C} \Delta
\Omega^{\rm C} = 0$.  Observations indicate that, when averaged over
many glitches, a few percent of the total spin-down is reversed in
glitches \citep{lg98}, corresponding to $I^{\rm SF}/I^{C} \approx
10^{-2}$.  It follows we can estimate the pre-glitch velocity lag by
\begin{equation}
\Delta v \approx \Delta \Omega^{\rm SF} R 
\approx \frac{I^{\rm C}}{I_{\rm SF}} \Delta \Omega^{\rm C} R ,
\end{equation}
where the quantity $\Delta \Omega^{C}$ is obtained directly from
glitch observations.  Parameterising with the Vela in mind:
\begin{equation}
\label{eq:v_glitch}
\Delta v \approx 7 \times 10^3 {\, \rm cm \, s}^{-1} \, 
\left(\frac{10^{-2}}{I^{\rm SF}/I}\right)
\left(\frac{\Delta \Omega^{\rm C}}{7 \times 10^{-5} {\, \rm Hz}}\right) .
\end{equation}
Comparing with equation (\ref{eq:v_critical_param}), we see that this
is smaller than the relative velocity needed to sustain our
non-precessing stars, for at least some part of the potential
parameter space (viz rapidly spinning stars with large deformations
and low pinned superfluid fractions).  However, it is likely that
(\ref{eq:v_glitch}) is too conservative an estimate on the critical
velocity for unpinning.  Firstly, the estimate assumes that the
velocity difference is reset to zero at the glitch.  If only some of
the velocity difference is relieved, a larger critical velocity would
be obtained.  Secondly, glitches presumably involve the most weakly
pinned superfluid (in a spinning down star, the weaker pinning sites
will break before the stronger ones), whereas it is the strongest
pinning locations that matter in building our non-precessional
solutions.  So, it is not clear if the relative velocity of equation
(\ref{eq:v_glitch}) limits the allowed vortex-superfluid velocities in
neutron stars.

Turn now to free precession.  The free precession period of a neutron
star in the absence of superfluid pinning is given approximately by
$P_{\rm fp} \sim P I/\Delta I$ \citep{ll76}, while the precession
period with pinning is approximately $P_{\rm fp} \sim P I/I^{\rm SF}$
\citep{shah77}.  Given that we expect $\Delta I \ll I^{\rm SF}$, it
follows that the precession period is much shorter in the case of
superfluid pinning than without pinning.  If follows that an
observation of free precession would yield an insight into the stellar
interior.  Unfortunately, free precession does not seem to be a common
phenomenon in the pulsar population.  The most convincing observation
is of approximately 500 day precession in PSR B1828-11 \citep{sls00},
but more recent observations of this and other objects may cast doubt
on the precessional interpretation (M. Kramer, personal
communication).  Nevertheless, if interpreted as precession, the long
500 day period implies that there is no significant pinned superfluid
component in the precessing star.  As discussed by \citet{link03,
  link06}, this may have significant repercussion for the nature of
superconductivity in the stellar core (but see \citet{gaj08, gaj09}
for a counterargument).  What does seem likely is that, if PSR
B1828-11 really is precessing, the precessional motion itself must
have set up a velocity field that has caused large scale unpinning
\citep{lc02}.  It follows that one can estimate the critical velocity
to up-pin even the strongest pinned vortices by examining the relative
flow in a precessing neutron star with superfluid pinning.  The
velocity field is of order \citep{gaj09}
\begin{equation}
\Delta v \le \Omega R \theta \frac{I^{\rm SF}}{I} ,
\end{equation}
where $\theta$, the `wobble angle', can be estimated from observations
\citep{ja01, le01}.  Parameterising in terms of PSR B1828-11:
\begin{equation}
\label{eq:v_precession}
\Delta v({\rm PSR \, B1828-11, \, precession}) 
\le 8.1 \times 10^3 {\, \rm cm \, s^{-1}}
\left(\frac{R}{10^6 \, \rm cm}\right)
\left(\frac{\theta}{3^\circ}\right)
\left(\frac{I^{\rm SF}/I}{10^{-2}}\right) .
\end{equation}
Comparing with equation (\ref{eq:v_critical_param}) we see that this
critical velocity is low enough to be problematic for some portion of
the parameter space of interest for our non-precessing solutions,
specifically very rapidly rotating stars with small superfluid
fractions.

Finally let us turn to theoretical estimates of the strength of
pinning.  In the case of pinning in the inner crust, \citet{link09}
recently examined the problem of how a vortex embedded in a nuclear
lattice responds to a superfluid flow.  Link solved the equation of
motion of a single vortex, including tension, Magnus and drag terms.
He found that when the relative vortex-superfluid velocity lay below a
critical value the vortex essentially remained bound to the nuclei,
i.e. the perfect pinning regime applied.  Using recent calculations of
the vortex-nucleus interaction energy \citep{dp06} Link estimated
\begin{equation}
  \label{eq:link_delta_v}
  \Delta v (\rm crust, \, theory) \approx 10^6{\rm-}10^7 \, \rm cm \, s^{-1} .
\end{equation}
Comparing (\ref{eq:link_delta_v}) and (\ref{eq:v_critical_param})
indicates that, by Link's calculation at least, our non-precessing
stars are safely in the regime of perfect pinning in the inner crust.
As discussed by \cite{link09}, this velocity difference is large
enough to allow the necessary accumulation of differential rotation to
explain giant glitches, but too large to be consistent with an
observation of slow free precession.  In the case of pinning in the
core, \citet{link03} estimates a maximum pinning force of $3 \times
10^{15} B_{12}^{1/2}$ dyn cm$^{-1}$, or in terms of a relative
velocity,
\begin{equation}
\Delta v (\rm core, \, theory) \approx 5 \times 10^3 {\, \rm cm \, s^{-1} \,}
\left(\frac{B}{10^{12} \, \rm G}\right)^{1/2}
\left(\frac{3 \times 10^{14} \rm \, g \, cm^{-3}}{\rho^{\rm SF}}\right) .
\end{equation}
Comparing with equation (\ref{eq:v_critical_param}) we see that, by
this estimate at least, the core vortex--fluxtube interaction may not
be strong enough to sustain pinning for non-precessing stars with high
spin frequencies and small superfluid fractions, and the problem is
more severe for weakly magnetised stars.

To sum up, the critical unpinning velocities implied by observations
suggest that unpinning may be a problem for our non-precessional
solutions, but only for very rapidly spinning stars with relatively
small pinned superfluid fractions.  Slower spinning stars, or stars
with larger pinned superfluid fractions, should be immune from
Magnus-force generated unpinning.  Theoretical estimates, on the other
hand, suggest that all non-precessional solutions should be safe from
unpinning in the inner crust.  Alternatively, one can note that an
observation of a non-precessing star radiating gravitationally at
$\Omega$ and $2\Omega$ would imply that the bounds on the critical
unpinning velocity from glitches and precession (equations
\ref{eq:v_glitch} and \ref{eq:v_precession}) are not reliable.

\subsection{Changes in shape}

Now let us examine the change in shape of the star in response to the
pinning forces.  We will consider the cases of elastic and magnetic
deformation separately.

Let us begin with a star deformed by a magnetic field of strength
$\sim B$.  The deformation $\Delta I$ will be sourced by a magnetic
stress of order $B^2$, and, to a rough approximation, the fractional
deformation $\Delta I/I$ will be of order of the ratio of
magnetostatic and gravitational binding energies \citep{jone02}:
\begin{equation}
\frac{\Delta I}{I} \sim \frac{B^2 R^3}{GM^2/R} .
\end{equation}
The Magnus force will perturb the star, resulting in a magnetic stress
perturbation of order $B \delta B$, and so a magnetic force per unit
volume $\sim B \delta B/R$.  This balances the Magnus force per unit
volume of $\rho^{\rm SF} R \Omega^2 \Delta I/I^{\rm SF}$.  Combining
these results, and simplifying using $I^{\rm SF} \sim \rho^{\rm SF}
R^5$ leads to
\begin{equation}
\frac{\delta B}{B} \sim \frac{I\Omega^2}{GM^2/R} .
\end{equation}
This is the ratio of the rotational kinetic energy to the
gravitational binding energy, and will be much less than unity for all
but the most rapidly rotating stars.  We can therefore conclude that,
in the case of a magnetically deformed star, the change in magnetic
field, and therefore shape, caused by the pinning forces is small.

A very similar calculation applies is we assume the pinning occurs in
a crustal shell of thickness $\Delta R$, shear modulus $\mu$, subject
to a strain $u$.  The deformation $\Delta I/I$ will be sourced by an
elastic stress of order $\mu u$, and, to a rough approximation,
$\Delta I/I$ will be of order of the ratio of Coulomb binding energy
to gravitational binding energy, multiplied by the strain
\citep{jone02}:
\begin{equation}
\frac{\Delta I}{I} \sim \frac{\mu R^2 \Delta R}{GM^2/R} u .
\end{equation}
The Magnus force will perturb the crust, resulting in an elastic
stress of order $\mu \delta u$, and so an elastic force per unit
volume of $\mu \delta u/\Delta R$.  This balances the Magnus force per
unit volume $\rho^{\rm SF} R \Omega^2 \Delta I/I^{\rm SF}$.  Combining
these results, and simplifying using $I^{\rm SF} \sim \rho^{\rm SF}
R^4 \Delta R$ leads to
\begin{equation}
\frac{\delta u}{u} \sim \frac{\Delta R}{R} \frac{I\Omega^2}{GM^2/R} .
\end{equation}
This is a factor $\Delta R/R$ smaller than the corresponding magnetic
result, showing that, in the elastic case, the pinning forces have
minimal impact on the strain distribution.

We can therefore conclude that pinning forces do not significantly
alter the shape of our star.  This means there is no danger of the
pinning inducing fracture, and also that the gravitational wave field
calculated previously is not significantly modified by Magnus
force-induced shape changes.

\subsection{Superfluid instabilities}

We can also ask whether or not a superfluid vortex instability might
be induced, of the form investigated recently by several authors
\citep{pmgo06, sac08}.  The instability occurs in stars consisting of
two fluid components: a superfluid neutron component, and a charged
component, coupled via the mutual friction force.  When the coupling
between superfluid vortices and the star's charged components is
sufficiently strong, a relative flow of superfluid along the vortex
array induces an inertial plane wave instability.  In particular,
\citet{gaj08, gaj09} showed that such an instability would be
triggered in a precessing star, if the wobble angle were sufficiently
large.  Could such an instability be triggered by the angular velocity
mismatch in our non-precessing star?

Taking the strong pinning limit of \citet{gaj09} (their equations (46)
and (47)) and writing the relative flow of superfluid along vortices,
$w$, as $w \approx \Omega R \Delta I/I_{\rm SF}$ as given in equation
(\ref{eq:v_critical_nonparam}), we find that there do indeed exist
unstable plane wave solutions; see appendix \ref{sect:app_tsi} for
details.  These occur for wavelengths smaller than a critical value:
$\lambda_{\rm crit} \approx \pi R \Delta I/I_{\rm SF}$:
\begin{equation}
\lambda^{\rm critical} \approx 30 {\, \rm cm \,}
\left(\frac{R}{10^6 \, \rm cm}\right)
\left(\frac{\Delta I/I}{10^{-7}}\right)
\left(\frac{10^{-2}}{I^{\rm SF}/I}\right) .
\end{equation}
This wavelength is longer than the inter-vortex separation in all
stars of interest \citep{saul89}:
\begin{equation}
\Delta x^{\rm vortex} \approx 10^{-2} {\, \rm cm \,} 
\left(\frac{\rm Hz}{f}\right)^{1/2} ,
\end{equation}
leaving a range of wavelengths where the instability could be
operative.  As shown in appendix \ref{sect:app_tsi}, the fastest
growing perturbation has a wavelength of approximately $\lambda^{\rm
  critical}/2$, and a corresponding growth time of
\begin{equation}
\tau \approx 0.1 {\, \rm s \,}
\left(\frac{100 \, \rm Hz}{f}\right)^2
\left(\frac{10^8 \, \rm K}{T}\right)^2
\left(\frac{x_p}{0.1}\right)
\left(\frac{10^6 \, \rm cm}{R}\right)^2
\left(\frac{I^{\rm SF}/I}{10^{-2}}\right)^2
\left(\frac{10^{-7}}{\Delta I/I}\right)^2 .
\end{equation}
In this equation $x_p$ denote the ratio of the proton to neutron
densities, and the temperature $T$ enters as the instability is
sensitive to the temperature dependent electron-electron shear
viscosity.  These results indicate that the superfluid instability may
be operative in stars of gravitational wave interest.  If so, the
instability could potentially unpin the vortices and, presumably,
eliminate the misalignment between superfluid and crustal principal
axis necessary for our wave generation mechanism, eliminating the
multiple harmonic gravitational wave emission.

However, \citet{vl08} have recently argued that the inclusion of
hydromagnetic forces acts to stabilise the instability.  They found
that, when such forces were included, the relative velocity needed for
instability was $(B B_{\rm cr}/\pi \rho^{\rm SF})^{1/2}$, where
$B_{\rm cr} \approx 10^{15}$ G is the critical magnetic field confined
to the flux tubes.  Combing with our relative velocity $w \approx
\Omega R \Delta I/I^{\rm SF}$ allows us to estimate the spin frequency
\emph{above} which this inertial wave instability would operate:
\begin{equation}
f \approx 3 \times 10^4 {\, \rm Hz \,}
\left(\frac{10^6 \, \rm cm}{R}\right)
\left(\frac{10^{-7}}{\Delta I/I}\right)
\left(\frac{I^{\rm SF}/I}{10^{-2}}\right)
\left(\frac{10^{14} \, \rm g \, cm^{-3}}{\rho^{\rm SF}}\right)^{1/2}
\left(\frac{B}{10^{12} \, \rm G}\right)^{1/2} .
\end{equation}
This is very high, indicating that the magnetohydrodynamic forces
might indeed be effective in suppressing the instability in all but
the most weakly magnetised stars.

To sum up, the multipole moments $Q_{21}$ and $Q_{22}$ that generate
our gravitational wave field will be bounded by the crust's finite
shear modulus and breaking strain, as described by \citet{ucb00}.
There are other possible failure mechanisms connected with the finite
strength of the pinning and a possible vortex instability.  Depending
upon the detailed (and poorly understood) physics of the amount and
location of the pinned superfluid and the pinning strength, these
failure mechanism could potentially prevent the creation of a steadily
rotating $Q_{21}$ mass quadrupole.  Conversely, observation of
gravitational radiation at the two frequencies $(\Omega, 2\Omega)$
would imply that these failure mechanism are not operative.

\section{Other frequency splitting mechanisms}
\label{sect:ofsm}

There exists (at least) two other ways in which an apparently steadily
spinning star might emit at both $\Omega$ and $2\Omega$, even without
a pinned superfluid component.  We will comment on these mechanisms
briefly, and argue that they are much less efficient than superfluid
pinning in producing the double frequency emission.

\subsection{Free precession of a biaxial star}

Consider a perfectly biaxial star, i.e.\ one with moment of inertia
tensor of the form
\begin{equation}
I_{ab} = I_0 \delta_{ab} + \Delta I n_a n_b .
\end{equation}
As is well known the free precession of such an object is very simple,
consisting of the superposition of two steady rotations: the symmetry
axis $n_a$ rotates about the fixed angular momentum vector with period
$\sim P$, with a superposed slow $(P_{\rm fp} \sim P I_0/\Delta I)$
rotation about the symmetry axis \citep{ll76}.  If the pulsar beam
lies exactly along $n_a$, and if the beam is exactly axisymmetric
about $n_a$, then the radio pulsations will be steady at period $P$,
with no precessional modulation \citep{ja01}.  The gravitational wave
emission is insensitive to the superimposed slow rotation, and has the
form given by equations
(\ref{eq:h_plus_biaxial})--(\ref{eq:h_cross_biaxial}), with
$I_{31}=\Delta I$ \citep{zs79, ja02}.  This is the sort of motion
modelled in \citet{gtt84}, \citet{bg96}, \citet{mp05}, and
\citet{vm09}, where a magnetic field sources the deformation $\Delta
I$; see section \ref{sect:discussion} for discussion.

However, this is clearly a very special case.  Firstly, it is unlikely
that the moment of inertia tensor is biaxial around the magnetic axis.
Crustal contributions are likely to be misaligned from the magnetic
ones, particularly if the crust retains any memory of a relaxed shape
from a previous higher rotation rate \citep{cul03}, and any
significant contribution to $I_{ab}$ not of the form $\Delta I n_a
n_b$ would make the inertia tensor triaxial, breaking the symmetry and
introducing detectable periodicities into the observed rotation
\citep{zs79, zimm80, vdb05}.  Secondly, pulsation profiles are known
to be non-axisymmetric, as evidenced by the complexity of pulse
profiles \citep{lg98}.  Finally, the precession motion would damp on
some dissipative timescale, so would not be long-lived \citep{swc99,
  cutl02}.  We therefore feel that such a motion is unlikely to be
common in the pulsar population.

\subsection{Electromagnetic torques}

Magnetised neutron stars are acted upon by an electromagnetic torque.
The torque consists of the familiar spin-down torque, scaling as
$\Omega^3$, and also a so-called anomalous torque, a factor $c/(\Omega
R)$ larger than the spin-down torque, scaling as $\Omega^2$
\citep{gold70}:
\begin{equation}
T_a = \frac{\alpha}{Rc^2} (\Omega^{\rm C}_d m_d) \epsilon_{abc} 
\Omega^{\rm C}_b m_c ,
\end{equation}
where $m_a$ is the star's dipole moment and $\alpha$ a factor of order
unity, which \citet{gold70} sets to unity, but \citet{gn85} correct to
$-1/5$.  This torque is orthogonal to $\Omega^{\rm C}_a$ and so does
no work on the star.  However, it can have an affect upon the rotation
\citep{gold70, mela00}.  To see this, look at the special case of a
star with a biaxial deformation $\Delta I^{\rm C}$ caused by crustal
strain and a biaxial deformation $\Delta I^{\rm B}$ caused by magnetic
fields:
\begin{equation}
I_{ab} = I_0 \delta_{ab} 
+ \Delta I^{\rm C} n^{\rm C}_a n^{\rm C}_b 
+ \Delta I^{\rm C} n^{\rm B}_a n^{\rm B}_b ,
\end{equation}
where the unit vectors $n^{\rm C}_a$ and $n^{\rm B}_a$ define the
symmetry axes of the crustal and magnetic deformations, respectively.
The Euler equation of motion in the presence of a torque $T_a$ is
\begin{equation}
\frac{dJ_a}{dt} + \epsilon_{abc} \Omega^{\rm C}_b J_c = T_a .
\end{equation}
Combining the above equations and selecting the non-precessional
solution by setting $dJ_a/dt = 0$ we obtain
\begin{equation}
\Delta I^{\rm C} (n^{\rm C}_d \Omega^{\rm C}_d) 
\epsilon_{abc} \Omega^{\rm C}_b n^{\rm C}_c +
\Delta I^{\rm C} (n^{\rm B}_d \Omega^{\rm C}_d) 
\epsilon_{abc} \Omega^{\rm C}_b n^{\rm B}_c
=
A (\Omega^{\rm C}_d n^{\rm B}_d) \epsilon_{abc} \Omega^{\rm C}_b n^{\rm B}_c ,
\end{equation}
where $m_a = m n^{\rm B}_a$ and
\begin{equation}
A = \frac{\alpha m^2}{Rc^2} .
\end{equation}
As noted by \citet{gold70}, this torque can be absorbed into the
moment of inertia tensor: 
\begin{equation}
  \epsilon_{abc} \Omega^{\rm C}_b [ \Delta I^{\rm C}
  (n^{\rm C}_d \Omega^{\rm C}_d) n^{\rm C}_c + (\Delta I^{\rm B}-A) (n^{\rm B}_d
  \Omega^{\rm C}_d) n^{\rm B}_c ] = 0 .
\end{equation}
It follows that the non-precessional solution is misaligned with the
principal axes of $I_{ab}$ by an angle of order
\begin{equation}
\delta \theta \sim \frac{A}{\Delta I} .
\end{equation}
This misalignment will create a non-axisymmetric mass distribution and
lead to gravitational wave emission.  That electromagnetic torques
have this effect was notes by \citet{swc99}, who considered the effect
of the non-anomalous spin-down torque; the principle is exactly the
same.

To gauge the significance of this, note that the magnetic moment $m$
is related to the (surface) magnetic field strength via $m \sim BR^3$,
and, the magnetic deformation scales as $\Delta I^{\rm B} \sim B^2
R^6/(GM)$, so that
\begin{equation}
A \sim \frac{R_{\rm Sch}}{R} \Delta I^{\rm B} ,
\end{equation}
where $R_{\rm Sch} \sim GM/(Rc^2)$ is of order the star's
Schwartschild radius.  If follows that the misalignment angle is of
order
\begin{equation}
  \theta \sim \frac{\Delta I^{\rm B}}{\Delta I} \frac{R_{\rm Sch}}{R} .
\end{equation}
It follows that this misalignment is very small unless the magnetic
field makes a significant contribution to the moment of inertia
tensor, and even in the case $\Delta I^{\rm B} \sim \Delta I^{\rm C}
\sim \Delta I$ the angle is suppressed by the factor $R_{\rm Sch}/R
\approx 0.2$.  We can therefore conclude that the anomalous torque is
not as effective as pinned superfluidity in modifying the
gravitational wave spectrum.

In fact, there is a further caveat to be attached to this.  The above
argument shows that there is a misalignment between the spin and
principal axes in a steadily spinning star.  If the gravitational
wave emission is then calculated by integration over the mass
distribution, emission at frequencies $\Omega$ and $2\Omega$ would be
found in the standard way.  However, one should presumably include the
contributions to the stress-energy tensor from the magnetic field
itself in this calculation.  One might worry that inclusion of this
additional component might eliminate the $\Omega$ harmonic.  This is
clearly an interesting issue, which we will address in a future study.

\section{Discussion}
\label{sect:discussion}

Most targeted gravitational wave searches to date have been motivated
by the possibility of the stars having `mountains', i.e. non-zero
values of the $Q_{22}$ mass multipole moment.  Such a deformed star,
rotating rigidly about a principal axis, would emit at exactly twice
the spin frequency.  Radiation at frequency $\Omega$ has received
relatively little attention.  In large part, this is because, in a
single component rigid star, the necessary $Q_{21}$ mass multipole
moment would require the star to precess, modulating the observed
timing of the radio pulses, and there has been little clear sign of
such modulation in the radio pulsations of gravitational wave
candidates.

In this paper we have argued that a neutron star containing a pinned
superfluid component can rotate in a steady non-precessional way,
radiating gravitationally at both $\Omega$ and $2\Omega$, without
there being any modulation in the observed radio pulsations.  Fundamentally, this is because the pinned superfluid effectively  acts as a gyroscope, `sewn' into the crust of the star.  This adds an extra piece to the angular momentum, allowing the system's total angular moment vector and rotation axis to coincide, even though no one of the crustal principal axes  lies along the rotation axis.  The
superfluid pinning is crucial, and the axis of pinning must not be
aligned with a principal axis of the moment of inertia tensor; if it
is, the non-precessional motion is about this axis, and only $2\Omega$ radiation is produced.  As we have noted, there
exist a number of mechanisms which could break the pinning, depending
upon the exact nature of how the superfluid vortices interact with the
rest of the star.

What does this mean for gravitational wave search strategies?
Clearly, this motivates carrying out gravitational wave searches where
data from two harmonics are combined to give a single detection
statistic, as described for biaxial precessing stars by \citet{jks98}.
The advantage of such a search over a single frequency search is that
it would capture the full radiated signal.  However, there is also a
disadvantage: the two-harmonic signal is specified by 10 parameters,
whereas the conventional single frequency search has only 7
parameters.  It is therefore likely that the signal-to-noise threshold
necessary to claim a detection would be larger in the two-harmonic
search, as the larger parameter space increases the chance of a false
detection.  Ultimately, just how useful the two-frequency search is
compared to the single frequency one will depend upon what sources
nature chooses to provide.  For a star that radiates only at twice the
rotation rate, the two frequency search simply degrades our
sensitivity to detection, without adding anything to the
signal-to-noise.  On the other hand, for a star that radiates
appreciably at both harmonics, the two frequency search would surely
be an advantage.  According to the analysis of section \ref{sect:gwa}, the two multipole moments contribute equally to the signal-to-noise ratio when $|Q_{21}| \approx 4 |Q_{22}|$ (neglecting the variation  in the detector noise with frequency).   In the extreme case of a targeted search, if the
star radiates only via the $Q_{21}$ multipole, the two frequency
search would be essential.  Clearly, some numerical experimentation to
quantify when you gain and when you loose would be useful.  One simple
possibility, motivated by the less efficient nature of gravitational
wave emission from the $Q_{21}$ moment, would be to search at a
single frequency, and only in the event of a detection look for
harmonic structure.

However, all of this begs the question: Why should the $Q_{21}$ mass
distribution be non-zero?  What astrophysical scenario might produce
such a state?  Unfortunately, this is a far more difficult problem to
address.  For stars spinning rapidly enough to be of gravitational
wave interest, the dominant force deforming neutron stars away from
spherical symmetry is rotation, which is necessarily axisymmetric and
can only contribute to the (non-radiative) $Q_{20}$ mass multipole.
It is therefore not obvious what might lead to a significant non-zero
value of $Q_{21}$.  However, exactly the same problem applies in
arguing for the existence of a $Q_{22}$ asymmetry.  So, by this
measure at least, there is no good argument to strongly favour either
one of the two sorts of gravitational wave producing asymmetries over
and above the other.

One possibility for producing asymmetry is a strong internal magnetic
field \citep{gtt84, bg96, hetal08, cfgp09, lj09}.  For instance, a
magnetic field, symmetric about an axis along $n^{\rm B}_a$ inclined
at an angle $\chi$ to the rotation axis, would produce a $\hat Q_{20}$
mass multipole when referred to a coordinate system with $\hat z$-axis
along direction $n^{\rm B}_a$, which, for $0 < \chi < \pi/2$,
translates into an $\chi$-dependent linear combination of $Q_{20},
Q_{21}, Q_{22}$ multipole moments, as given by the transformation
properties of spherical harmonics under rotation:
\begin{equation}
\label{eq:Y_transformation}
\hat Y_{20} = 
-\frac{3}{4\sqrt 6} \sin^2\chi (Y_{22}+Y_{2 -2})
-\frac{3}{\sqrt 6}i \sin\chi \cos\chi (Y_{21}+Y_{2 -1})
+\frac{1}{2} (2\cos^2\chi - \sin^2\chi) Y_{20} .
\end{equation}
However, as is well known, the strong field neutron stars (the
magnetars) all rotate too slowly to be of gravitational wave interest,
while the more rapidly spinning stars (the young pulsars, the low-mass
X-ray binaries (LMXBs), and the millisecond pulsars) have relatively
weak external magnetic fields.  Clearly, only if the internal magnetic
field far exceeds the external one would magnetic deformations be of
interest.

Most of the known pulsars do indeed seem to be `inclined rotators',
i.e. have magnetic fields that are neither aligned with ($\chi = 0$)
nor orthogonal to ($\chi = \pi/2$) the rotation axis.  What does this
mean for their gravitational wave emission?  The answer to this
depends upon whether or not there is a pinned superfluid component.
In the absence of such a component, there are two possibilities.
Firstly, if the magnetic deformation were the only deformation, the
star would precess, and either align or go orthogonal on some short
timescale, leading to no gravitational wave emission or emission at
only $2\Omega$, respectively \citep{cutl02}.  Secondly, and more
plausibly, there would be a corresponding crustal contribution to the
inertia tensor, such that the inclined rotator simply rotates about a
principal axis of the total inertia tensor, radiating at $2\Omega$
only.  (As discussed in detail by \citet{wass03}, for a given
inclination angle $\chi$ and a given magnetic deformation, this would
require a sufficiently large crustal deformation).  However, if there
is a pinned superfluid component, then the analysis of this paper
applies, the motion is non-precessional, and emission at both $\Omega$
and $2\Omega$ is possible, depending upon the orientation of the
pinned superfluid with respect to the principal axes of the total
inertia tensor.

One particular scenario where magnetic fields might prove to be
important is in accreting systems, if magnetic burial of the field
takes place, resulting in stars with fast spins, strong internal
fields and weak external ones.  Such a scenario has been modelled in
detail by \citet{mp05} and \citet{vm09}, who showed that accretion of
$\Delta M \sim 2 \times 10^{-3} M_\odot$ was sufficient to generate a
mass quadrupole of size $\Delta I/I \approx 2 \times 10^{-5}$, biaxial
about the magnetic axis.  As noted by the above authors, this would
generate a precessing star, radiating gravitational waves at the
$\Omega, 2\Omega$ harmonics.

However, in the absence of superfluid pinning, there is an objection
to this scenario.  The precessional motion would damp on a timescale
much shorter than the timescale on which the magnetic mountain is
built: the quadrupole is generated on a timescale related to the
accretion rate $\dot M$, roughly $\Delta M/\dot M \sim 10^6$ years for
a typical LMXB, while the precession period, in the absence of pinned
superfluidity, is of order $P/\epsilon \sim 10^2$ seconds.  It follows
that the quality factor of the precession would have to be
unrealistically large to allow the precession, and therefore the
radiation at frequency $\Omega$, to survive \citep{ja01, cutl02}.
This changes completely if a pinned superfluid is added to the model.
In this case, the rotation remains close to the superfluid pinning
axis, and there is no precession, allowing the inclined biaxial
magnetic mountain to radiate continuously at the two frequencies.  So,
unless there exists some mechanism to maintain the precession, the
superfluid pinning model of this paper is essential to realise the
long-lived multiple frequency gravitational wave generation described
by \citet{mp05} and \citet{vm09}.

The solid crust is the other possible source of the deformation.  As
was computed in detail by \citet{cul03}, the crust in a pulsar is
likely to have a $Q_{20}$ density perturbation sourced by strains
generated by the crust retaining a memory of a relaxed state at a
higher rotation rate.  If such a star is kicked, precession occurs,
generating the familiar gravitational wave field.  However, what would
generate non-zero $Q_{21}$ (or, for that matter, $Q_{22}$) for our
non-precessing stars?  One possibility is that some of the glitches in
young stars correspond to starquakes, i.e. sudden cracking events, and
that these don't occur in completely axisymmetric way.

In fact, a model of symmetry breaking in starquakes has been developed
by \citet{fle00}, who were attempting to explain the permanent
increases in spin down rate seen after glitches in the Crab pulsar.
Their suggestion was that the crust cracks in a non-axisymmetric way,
so that, immediately after the glitch, the spin and principal axes are
no longer coincident.  They argue that a brief period of precession
would follow, which, when damped, would leave the star with a
permanently increased angle between the magnetic and spin axes.
Crucially, to account for the systematic increase of this angle,
Franco et al.\ argued that the magnetic field itself breaks the
axisymmetry in the strain build up, and favours quakes along fault
lines which tend to increase the spin axis--magnetic dipole axis
angle.  Clearly, such non-axisymmetric faulting could, in our pinned
superfluid scenario, contribute to building non-zero values of the
$Q_{21}$ and $Q_{22}$ multipoles.  However, for this to work, the
starquake would have to occur without significantly unpinning the
superfluid, or at least the pinning would have to reform on a
timescale short compared to the timescale on which the precession of
the star is damped, or else the system would relax to rotation about a
principal axis.

The neutron star crust tectonics model of Ruderman and collaborators
provides another possible mechanism of symmetry breaking
\citep{rude76, rude91a, rude91b}.  This mechanism makes use of both
the magnetic field and the core superfluid.  The idea is that as a
star spins down and its core superfluid vortices migrate outwards,
they interact strongly with the magnetic flux tubes which would thread
a type II superconducting core.  This places a large strain on the
crust.  The exact deformation produced (i.e. the sizes of $Q_{21}$ and
$Q_{22}$) would require a detailed calculation, and would depend upon
the poorly understood interaction between the vortices and flux tubes
\citep{sa09}, but Ruderman estimates that the corresponding stresses
are large--large enough to shear the crust, and so could be a source
of non-axisymmetry.

To sum up, we have argued that pinned superfluidity allows a neutron
star of spin frequency $\Omega$ to emit steady gravitational wave
signals at both $\Omega$ and $2\Omega$, without there being any
stellar precession.  This motivates gravitational wave searches at
both harmonics, even when targeting known pulsars with very smooth
timing profiles.  The non-detection of such radiation would imply that
the necessary gravitational wave generating quadrupole moments aren't
created in real stars, and/or the pinning isn't strong enough to
support them.  More interestingly, the observation of radiation at
both harmonics from a non-precessing star would provide evidence in
favour of pinned superfluidity within the star.

\section*{Acknowledgements}

The author would like to thank Kostas Glampedakis for stimulating
discussions on several of the issues discussed in this paper, and also
members of the LSC/Virgo continuous wave group.  This work was
supported by PPARC/STFC via grant number PP/E001025/1.  The author
also acknowledges support from COMPSTAR, an ESF Research Networking
Programme.

\appendix

\section{The coupling torque}
\label{sect:app_tct}

In our model, the superfluid component does not rotate steadily---its
angular velocity vector $\Omega^{\rm SF}_a$ rotates about the fixed
crustal angular velocity vector $\Omega^{\rm C}_a$.  In this appendix
we will show that the torque required to support this motion is
supplied by the Magnus force which couples the superfluid and crustal
components via the superfluid vortices.

The superfluid is acted on by a Magnus force \citep{shah77}
\begin{equation}
f_a = \rho^{\rm SF} \epsilon_{abc} \omega_b (v^{\rm SF}_c - v^{\rm C}_c) ,
\end{equation}
where $v^{\rm SF}_a = \epsilon_{abc} \Omega^{\rm SF}_b x_c$ is the
superfluid velocity, $v^{\rm C}_a = \epsilon_{abc} \Omega^{\rm C}_b
x_c$ the crust velocity, and $\omega_a = 2\Omega^{\rm SF}_a$ the
superfluid vorticity.  We want to see what sort of torque this
produces on the crust:
\begin{equation}
T_a = \int_V \epsilon_{abc} x_b f_c \, dV .
\end{equation}
Combining the above equations and simplifying we find
\begin{equation}
\label{eq:torque}
T_a = 2 \epsilon_{abc} \int_V \rho^{\rm SF} x_b x_d \, dV \, 
\Omega^{\rm SF}_d  (\Omega^{\rm SF}_c - \Omega^{\rm C}_c) .
\end{equation}
In spherical symmetry (which we assume for the superfluid) the
integral simplifies:
\begin{equation}
\int_V \rho^{\rm SF} x_b x_d \, dV = 
\delta_{bd} \int_V \rho^{\rm SF} x^2 \, dV ,
\end{equation}
while the moment of inertia tensor is, by definition,
\begin{equation}
I^{\rm SF}_{ab} = \int_v \rho^{\rm SF} (r^2 \delta_{ab} - x_a x_b) \, dV ,
\end{equation}
where $r^2 = x^2 + y^2 + z^2$.  Picking out a single component:
\begin{equation}
I^{\rm SF}_{xx} = \int_V \rho^{\rm SF} (y^2+z^2) \, dV 
= 2 \int_V \rho^{\rm SF} x^2 \, dV ,
\end{equation}
and so
\begin{equation}
\int_V \rho^{\rm SF} x_b x_d \, dV = \delta_{bd} I^{\rm SF}/2 ,
\end{equation}
where $I^{\rm SF} \equiv I^{\rm SF}_{xx} = I^{\rm SF}_{yy} = I^{\rm
  SF}_{zz}$ is the moment of inertia of the spherical superfluid.
Inserting this into (\ref{eq:torque}) and writing in terms of the
superfluid angular momentum $J^{\rm SF}_a = I^{\rm SF} \Omega_a$ gives
\begin{equation}
T_a = \epsilon_{abc} \Omega^{\rm C}_b J^{\rm SF}_c .
\end{equation}
This is exactly the torque required to spin a superfluid of angular
momentum $J^{\rm SF}_a$ at a rate $\Omega^{\rm C}_a$, confirming that
the Magnus force does indeed supply the necessary torque to provide a
self-consistent solution.

\section{The superfluid instability}
\label{sect:app_tsi}

The short wavelength superfluid instability is described in detail in
\citet{gaj09}.  The complex plane wave frequencies that apply in the
limit of strong vortex drag are given by equations (46) and (47) of
their paper.  The unstable mode has frequency
\begin{equation}
\sigma \approx \Omega^{\rm SF} 
\left(1 - \frac{1}{x_p}\right) + kw + \frac{i}{2} \nu k^2
+\frac{1}{x_p}\left[
-\frac{1}{4}(\nu k^2 x_p)^2 + i x_p(x_p-1)\Omega^{\rm SF} \nu k^2
+\Omega_{\rm SF}^2 (1+x_p)^2 - 2\Omega^{\rm SF} k w x_p\right]^{1/2} ,
\end{equation}
where $x_{\rm p}$ is the ratio of proton to neutron densities, $k$ the
wave number, and $\nu$ the electron-electron shear viscosity.  Taking
the limit of small wavelength and retaining the dominant terms in the
imaginary part of the mode frequency we find:
\begin{equation}
\Im(\sigma) \approx \frac{2\Omega^{\rm SF}}{x_p \nu}
\left[\frac{2\Omega^{\rm SF}}{k^2} - \frac{w}{k}\right] .
\end{equation}
Inserting the relation $w \approx R \Omega^{\rm SF} \Delta I/I^{\rm
  SF}$ gives
\begin{equation}
\label{eq:im_sigma}
\Im(\sigma) \approx \frac{2\Omega^2}{\nu x_p}
\left[\frac{2}{k^2} - \frac{R}{k}\frac{\Delta I}{I^{\rm SF}}\right] .
\end{equation}
The mode is unstable for $\Im(\sigma) < 0$, so the plane waves are
unstable for wavelength less than a critical value at which
$\Im(\sigma) = 0$:
\begin{equation}
\lambda^{\rm critical} = \pi R \frac{\Delta I}{I^{\rm SF}} .
\end{equation}
Further analysis of (\ref{eq:im_sigma}) shows that the fastest growing
mode has a wavelength $\lambda^{\rm critical}/2$.  Substituting this
wavelength into (\ref{eq:im_sigma}) and inverting gives the mode
growth timescale:
\begin{equation}
\tau = \frac{4\nu x_p}{\Omega^2 R^2} 
\left(\frac{I^{\rm SF}}{\Delta I}\right)^2 .
\end{equation}
This gives a good estimate of the growth time of the most unstable
mode.


\label{lastpage}

\end{document}